\documentclass[prd, aps, floatfix, 10pt, twocolumn, superscriptaddress, preprintnumbers,nofootinbib]{revtex4-2}

\usepackage{empheq}
\usepackage{textcomp}
\usepackage[latin1]{inputenc}                    
\usepackage{graphicx}                            
\usepackage{epsfig}
\usepackage[export]{adjustbox}
\usepackage{mathrsfs}
\usepackage{mathtools}
\usepackage{soul}
\usepackage{epsf}
\usepackage{latexsym}                            
\usepackage{amsfonts}                            
\usepackage{amssymb}                             
\usepackage{amsmath}                             
\usepackage{slashed}
\usepackage[mathscr]{eucal}                      
\usepackage{dcolumn}                             
\usepackage{multirow}                               
\usepackage{bm}                                  
\usepackage[usenames]{color}
\usepackage[dvipsnames]{xcolor}

\definecolor{lcolor}{rgb}{0.5,0,0}
\definecolor{citcolor}{rgb}{0,0.3,0.0}

\usepackage[breaklinks,colorlinks,urlcolor=blue,citecolor=citcolor,linkcolor=lcolor]{hyperref}

\usepackage{color}
\usepackage{empheq}

\usepackage{siunitx}

\graphicspath{{.}{Graphics/}}                    
\newcommand{\msbar}{\overline{\mbox{\rm MS}}}  

\newcommand{\be}{\begin{equation}}
\newcommand{\ee}{\end{equation}}
\newcommand{\bea}{\begin{eqnarray}}
\newcommand{\eea}{\end{eqnarray}}

\def \3{\ss }

\newcommand{\beqn}{\begin{eqnarray}}
\newcommand{\eeqn}{\end{eqnarray}}

\usepackage{physics}

\newcommand{\nc}{N_c}
\newcommand{\xt}{\mathbf{x}}
\newcommand{\yt}{\mathbf{y}}
\newcommand{\rt}{\mathbf{r}}
\newcommand{\bt}{\mathbf{b}}
\newcommand{\kt}{\mathbf{k}}
\newcommand{\pt}{\mathbf{p}}

\newcommand{\qt}{\mathbf{q}}
\newcommand{\Deltat}{\boldsymbol{\Delta}}

\newcommand{\aem}{\alpha_\text{em}}
\newcommand{\as}{\alpha_s}
\newcommand{\xbj}{x_\text{Bj}}

\newcommand{\PS}{\text{PS}}

\newcommand{\li}{\text{Li}}

\begin{document}
\title{Matching collinear factorization with color-glass condensate for inclusive and exclusive deep inelastic scattering}

\author{Shohini Bhattacharya}
\email{shohinib@uconn.edu}
\affiliation{Department of Physics, University of Connecticut, Storrs, CT 06269, USA}

\author{Chuan-Qi He}
\email{legend\_he@m.scnu.edu.cn}
\affiliation{State Key Laboratory of Nuclear Physics and Technology, Institute of Quantum Matter, South China Normal University, Guangzhou 510006, China}

\author{Zhong-Bo Kang}
\email{zkang@physics.ucla.edu}
\affiliation{Department of Physics and Astronomy, University of California, Los Angeles, CA 90095, USA}
\affiliation{Mani L. Bhaumik Institute for Theoretical Physics, University of California, Los Angeles, CA 90095, USA}
\affiliation{Center for Frontiers in Nuclear Science, Stony Brook University, Stony Brook, NY 11794, USA}

\author{Diego Padilla}
\email{dpadi022@g.ucla.edu}
\affiliation{Department of Physics and Astronomy, University of California, Los Angeles, CA 90095, USA}
\affiliation{Mani L. Bhaumik Institute for Theoretical Physics, University of California, Los Angeles, CA 90095, USA}

\author{Jani Penttala}
\email{janipenttala@physics.ucla.edu}
\affiliation{Department of Physics and Astronomy, University of California, Los Angeles, CA 90095, USA}
\affiliation{Mani L. Bhaumik Institute for Theoretical Physics, University of California, Los Angeles, CA 90095, USA}

\begin{abstract}
Collinear factorization and color-glass condensate (CGC) effective field theory are generally treated as separate approaches for calculating scattering amplitudes, valid in different kinematic regimes.
For deep inelastic scattering at high photon virtuality and high center-of-mass energy, however, both of these approaches should be applicable.
By expressing collinear parton distributions and generalized parton distributions in the shockwave approximation, we show that the resulting collinear-factorization amplitudes exactly reproduce the large-$Q^2$ expansion of CGC amplitudes for inclusive deep inelastic scattering, deeply virtual Compton scattering, and deeply virtual meson production. The matching holds directly at the amplitude level and includes both logarithmically enhanced and finite contributions. Our results establish the consistency between collinear factorization and the CGC in their common region of validity, clarify the origin of large momentum logarithms within the CGC framework, and provide a path toward combining high-energy and collinear evolution in a unified description of hadronic structure at small $x$ and large momentum scales.
\end{abstract}

\maketitle

\section{Introduction}
\label{sec:introduction}

Perturbative methods for describing scattering processes involving initial hadrons require the use of factorization theorems that divide the process into perturbative and nonperturbative parts.
For processes with large momentum scales, such as inclusive deep inelastic scattering (DIS) with a large photon virtuality $Q^2$, the process is described within collinear factorization that encodes the nonperturbative information about the target hadron into parton distribution functions (PDFs)~\cite{Collins:2011zzd}.
In a different kinematic regime, corresponding to a high center-of-mass energy for the photon--target system, high-energy factorization provides a description of the target in terms of Wilson lines and their correlators. These Wilson-line correlators can then be understood as fluctuations of the target's classical color field with the color-glass condensate (CGC) effective field theory~\cite{Iancu:2003xm}.

While these two complementary approaches for describing hadrons have traditionally been treated as separate---especially in the context of nonlinear effects of QCD and gluon saturation~\cite{Morreale:2021pnn,Armesto:2022mxy}---they should yield matching results when both the momentum scale and the scattering energy are high.
Moreover, in this shared region of validity, the nonperturbative components of the different approaches have to be connected.
Indeed, matching between parton distributions and the CGC Wilson lines is now well understood at leading order (LO) in the eikonal approximation, as has been demonstrated in Refs.~\cite{Baier:1996sk,Mueller:1999wm,Mueller:2001fv,Hatta:2016dxp,Hatta:2017cte,Dominguez:2011wm,Marquet:2009ca,Hatta:2016dxp,Boussarie:2018zwg,Hatta:2022lzj,Hauksson:2024bvv,Xiao:2017yya,Zhou:2018lfq,Kovchegov:2025yyl,Kovchegov:2026gwb,Bhattacharya:2025fnz,Benic:2026idy}\footnote{Note that in our previous work Ref.~\cite{Bhattacharya:2025fnz} we overlooked some distributions that do not vanish in the eikonal limit. These have been considered in Ref.~\cite{Benic:2026idy}.}, along with recent works including subeikonal corrections~\cite{Altinoluk:2024tyx,Altinoluk:2024zom,Altinoluk:2025ang,Altinoluk:2025ivn,Fu:2023jqv,Fu:2024sba,Chirilli:2026pkv}.
Additionally, the matching has been studied for transverse-momentum-dependent (TMD) factorization~\cite{Xiao:2017yya,Taels:2022tza,Caucal:2022ulg,Caucal:2023fsf,Caucal:2023nci,Caucal:2024cdq,Caucal:2024nsb,Caucal:2025qjg,Caucal:2025xxh,Caucal:2024bae,Caucal:2025mth} and collinear factorization~\cite{Stasto:2014sea,Altinoluk:2014eka,Watanabe:2015tja,Hatta:2017cte,Boussarie:2023xun} at the level of the cross section.
However, what is still missing from the literature is showing that the obtained parton distributions can be directly used to calculate the high-energy cross section  in the collinear limit.
In this work, we aim to fix this by showing the consistency for various different processes in DIS.

Specifically, we wish to demonstrate that collinear factorization in the high-energy limit (or ``small $x$'') and CGC with large photon virtuality yield the same results for scattering processes.
For collinear factorization, we will do this by writing the parton distributions in the shockwave approximation for the target, where the nonperturbative degrees of freedom correspond to the CGC Wilson lines.
This approximation is valid in the high-energy limit, with the relevant parton distributions calculated in Ref.~\cite{Bhattacharya:2025fnz}.
For CGC, we will expand the scattering amplitudes in the limit of large photon virtuality $Q^2$, keeping only the most important contribution in the twist expansion.
The equivalence of the two results can then be understood as the commutativity of the high-energy and large-$Q^2$ limits.
By considering inclusive DIS, deeply virtual Compton scattering (DVCS), and deeply virtual meson production (DVMP), we will show this for a range of processes that cover both collinear PDFs and generalized parton distributions (GPDs), with both quark and gluon channels being important.
This procedure of directly using the parton distributions in the shockwave limit marks the difference to the previous works bridging collinear and high-energy factorization in DIS~\cite{Hatta:2017cte,Boussarie:2023xun}.

The article is organized as follows.
First, we recap high-energy factorization used in the CGC framework in Sec.~\ref{sec:amplitude} to clarify the connection between scattering amplitudes in the CGC and collinear frameworks.
In Sec.~\ref{sec:PDF}, we list the parton distributions that will be used in this work for completeness.
In the remaining sections, we show the matching between collinear factorization and CGC for DVMP (Sec.~\ref{sec:DVMP}), inclusive DIS (Sec.~\ref{sec:DIS}), and DVCS (Sec.~\ref{sec:DVCS}), before summarizing our results in Sec.~\ref{sec:conclusion}.

\section{Scattering amplitude in high-energy factorization}
\label{sec:amplitude}

To compare scattering amplitudes in the collinear factorization and in the CGC, let us briefly discuss the overall framework for calculating scattering amplitudes in the CGC approach.
The convenient framework for this is high-energy factorization, which separates the eikonal scattering off the target from the rest of the process.

To see this explicitly,
let us consider the subprocess $\gamma^* + p \to X + p'$ in DIS that is of interest in this work.
In the shockwave limit, we can write the scattering matrix as
\begin{equation}
\label{eq:S-matrix_eikonal}
\begin{split}
       &\mel{X p'}{\hat S}{\gamma^* p}
       = 2 \sqrt{p^+ p'^+}
       \int \dd{b^-} \dd[2]{\bt}
       \\
       &
       \times
       e^{- i (\pt' - \pt) \vdot \bt + i (p'^+ - p^+) b^-}
       \mel{X}{\expval*{\hat S}_{b}}{\gamma^*},
\end{split}
\end{equation}
where we have used the eikonal approximation~\cite{Bhattacharya:2025fnz}
\begin{equation}
\begin{split}
 \mel{p'}{\mathcal{O}}{p} =& 2 \sqrt{p^+ p'^+ }
 \int \dd{b^-} \dd[2]{\bt}
 \\
 & \times
 e^{-i (\pt' - \pt) \vdot \bt + i (p'^+ - p^+) b^- }
 \expval{\mathcal{O}}_b.
\end{split}
\end{equation}
Here $\mathcal{O}$ is an operator
and $\expval{\ldots}_{b}$ denotes the CGC average with the target at the coordinate $b = (\bt,b^-)$.
We are working in the coordinate frame where the incoming proton has a large plus-momentum, and the momenta of the incoming and outgoing protons are denoted by $p$ and $p'$, with bolded letters ($\pt$, $\pt'$) corresponding to transverse indices.
Specifically, we are working in the asymmetric frame where the momenta of the incoming and outgoing proton are given by:
\begin{align}
    p &= \qty([1+\xi]P^+, \frac{1}{2 [1+\xi]P^+} \qty[M^2 + \frac{1}{4} \Deltat^2], -\frac{1}{2}\Deltat),
    \\
    p' &= \qty([1-\xi]P^+, \frac{1}{2 [1-\xi]P^+} \qty[M^2 + \frac{1}{4} \Deltat^2], \frac{1}{2}\Deltat),
\end{align}
where $P = \frac{1}{2}(p + p')$, $\Delta = p' - p$, and $M$ is the proton mass.
We have also defined the skewness variable
\begin{equation}
    \xi = -\frac{\Delta^+}{P^+},
\end{equation}
and we note that the Mandelstam $t$-variable is given in terms of the momentum transfer $\Delta$ as
\begin{equation}
    t = (p' -p)^2 =  \Delta^2
    = -\frac{1}{1-\xi^2}\qty[ \Deltat^2+ 4 \xi^2 M^2].
\end{equation}

As the interaction with the target is now localized at $b^-$ in Eq.~\eqref{eq:S-matrix_eikonal}, it is convenient to work in light-cone perturbation theory~\cite{Kogut:1969xa,Bjorken:1970ah,Brodsky:1997de} with $x^-$ serving as the time coordinate.
The scattering operator is given by
\begin{equation}
    \hat S = U(x^-=+\infty, x^- = -\infty),
\end{equation}
where $U$ denotes the time-evolution operator in the interaction picture.
The interaction with the target only appears instantaneously at $b^-$, allowing us to write
\begin{equation}
  \expval*{\hat S}_{b} = 
  U(+\infty, b^-) \hat  U_p(b^-,\bt) U(b^-,-\infty).
\end{equation}
where $\hat U_p$ is the operator that contains the information about the scattering with the target.
We can now shift the target to the origin by using
\begin{equation}
    \mathcal{O}(x) = e^{i y \vdot \hat P}\mathcal{O}(x-y)
     e^{-i y \vdot \hat P}
\end{equation}
such that
\begin{equation}
      \mel{X}{\expval*{\hat S}_{b}}{\gamma^*}
      =
      e^{i b^- (p_X^+ - p_\gamma^+) - i \bt \vdot (\pt_X - \pt_\gamma)} \mel{X}{\expval*{\hat S}}{\gamma^*},
\end{equation}
where $\expval*{\hat S}\equiv\expval*{\hat S}_{b=0}$,
and thus
\begin{equation}
\begin{split}
       &\mel{X p'}{\hat S}{\gamma^* p}
       =  
       \mel{X}{\expval*{\hat S}}{\gamma^*}
        \times 2  (2\pi)^3 \sqrt{p^+ p'^+}
       \\
       &
     \times  \delta( p_\gamma^+ + p^+ - p_X^+ - p'^+ )
        \delta^{(2)}( \pt_\gamma + \pt - \pt_X - \pt' )
       .
\end{split}
\end{equation}

We can proceed further by introducing a complete set of Fock states $1 = \sum_n \int \dd{\PS_n} \ketbra{n}$
that are eigenstates of the free Hamiltonian $\hat P_0^+$ in the interaction picture. Here, we have used the shorthand notation $\ket{n} \equiv \ket{n(p_i^-, \pt_i)}$, where $i$ labels the particles in the Fock state $n$.
This allows us to separate $\hat U_p$ from the rest of the process:
\begin{equation}
\begin{split}
       &\mel{X p'}{\hat S}{\gamma^* p}
       =2  (2\pi)^3 \sqrt{p^+ p'^+}
       \\
       &
     \times  \delta( p_\gamma^+ + p^+ - p_X^+ - p'^+ )
        \delta^{(2)}( \pt_\gamma + \pt - \pt_X - \pt' )
        \\
        &
       \times
        \sum_{n n'} 
        \int \dd{[\PS]_n}\dd{[\PS]_{n'}}
\\
&
        \times
        \bra{X} U(+\infty, 0)\ket{n'}
        \mel{n'}{\hat U_p}{n}
        \bra{n}U(0, -\infty)\ket{\gamma^*}
      ,
\end{split}
\end{equation}
where the phase-space measure is given by
\begin{equation}
    \dd{\PS_n} = \prod_{i \in n}
    \frac{\dd[2]{\pt_i} \dd{p_i^-}}{(2\pi)^3 2 p_i^-}.
\end{equation}

The matrix elements before and after the shockwave can be written in terms of the light-cone wave functions $\Psi$~\cite{Beuf:2024msh,Penttala:2023udn}:
\begin{equation}
\begin{split}
        &\bra{n}U(0, -\infty)\ket{\gamma^*}
       \\
        =& 2 p_\gamma^- (2\pi)^3 \delta(p_n^- - p_\gamma^-) \delta^{(2)}(\pt_n - \pt_\gamma)
        \Psi^{\gamma^* \to n}_\text{in},
\end{split}
\end{equation}
\begin{equation}
        \begin{split}
     & \bra{X} U(+\infty, 0)\ket{n'}
     \\
        =& 2 p_{X}^- (2\pi)^3 \delta(p_{n'}^- - p_{X}^-) \delta^{(2)}(\pt_n' - \pt_X)
        \qty(\Psi^{X \to n'}_\text{out})^*.
        \end{split}
\end{equation}
Here, we distinguish the wave functions for the incoming and outgoing states as they differ in terms of the $i \varepsilon$ prescription in the Feynman propagators~\cite{Bjorken:1970ah,Beuf:2024msh,Penttala:2023udn}.

For the scattering off the shockwave, we note that the scattering is diagonal in terms of the transverse coordinates and does not depend on  the plus or minus coordinates~\cite{Balitsky:1995ub}.
It is thus useful to Fourier transform the transverse momenta of the state $n$ to the coordinate space as:
\begin{equation}
    \ket{n(p_i^-, \pt_i)}
    = 
    \int \prod_{i \in n}
    \qty[\dd[2]{\xt_i} e^{i \pt_i \vdot \xt_i} ]
    \ket{n(p_i^-, \xt_i)}
\end{equation}
with the normalization
\begin{equation}
\begin{split}
    &\bra{n(p'^-_i, \xt'_i)}\ket{n(p^-_i, \xt_i)}\\
    =& \prod_{i \in n} \qty[ 2p_i^- (2\pi) \delta(p_i^- - p_i'^-) \delta^{(2)}(\xt_i - \xt'_i) ].
\end{split}
\end{equation}
Then for each particle $i$ in the state $n$ we have
\begin{equation}
    \begin{split}
    &\mel{i'}{\hat U_p}{i}
    = 
    \int \dd[2]{\xt_i} \dd[2]{\xt'_i }
     e^{i \pt_i \vdot \xt_i -i \pt'_i \vdot \xt'_i} 
     \\
     &\times
    \mel{i'(p_i'^-, \xt'_i)}{U(\xt_i)}{i(p_i^-, \xt_i)}
    \\
    =&    
    2 p_i^- (2\pi) \delta(p_i^- -p_i'^-)
    \int \dd[2]{\xt_i} 
     e^{i (\pt_i-\pt_i') \vdot \xt_i} 
 U_i(\xt_i),
    \end{split}
\end{equation}
where $U_i$ is a color matrix in the representation of the particle $i$, and we have suppressed the color indices for clarity.
Each particle $i$ interacts with the target independently, such that the matrix element for the whole Fock state $n$ is given by
\begin{equation}
\begin{split}
    &\mel{n'}{\hat U_p}{n}
    \\
    = &
    \prod_{i \in n}
    \qty[
       2 p_i^- (2\pi) \delta(p_i^- -p_i'^-)
    \int \dd[2]{\xt_i} 
     e^{i (\pt_i-\pt_i') \vdot \xt_i} 
 U_i(\xt_i)
    ].
\end{split}
\end{equation}
We note that the dependence on the transverse momenta is only in the exponentials, and hence we can integrate over them by defining the mixed-space light-cone wave functions:
\begin{equation}
   \widetilde  \Psi^{m \to n}
   = \int \prod_{i \in n} \qty[
   \frac{\dd[2]{\pt_i}}{(2\pi)^2} e^{i \pt_i \vdot \xt_i}
   ] 
   \
  (2\pi)^2 \delta^{(2)}(\pt_m - \pt_n)
   \Psi^{m \to n}.
\end{equation}
The scattering matrix can now be written as
\begin{equation}
\label{eq:shockwave_scattering}
\begin{split}
       &\mel{X p'}{\hat S}{\gamma^* p}
       =(2\pi)^4\delta^{(4)}( p_\gamma + p - p_X - p' ) 
       \\
        &
        \times
        4  p^-_\gamma  \sqrt{p^+ p'^+}
        \sum_{n} 
        \int \dd{[\widetilde \PS]_n}
        2 p_\gamma^- (2\pi) \delta(p_\gamma^- - p_n^-)
\\
&
        \times
\widetilde \Psi_\text{in}^{\gamma^* \to n}
S^{(n)}
\qty(\widetilde \Psi_\text{out}^{X \to n})^*
      ,
\end{split}
\end{equation}
where $S^{(n)} = \prod_{i \in n} U_i(\xt_i)$
and the mixed-space phase-space measure is
\begin{equation}
    \dd{\widetilde \PS_n} = \prod_{i \in n}
    \frac{\dd[2]{\xt_i} \dd{p_i^-}}{(2\pi) 2 p_i^-}.
\end{equation}
We have also approximated
\begin{equation}
    \delta( p^-_\gamma + p^- - p_X^- - p'^- )
    \approx
    \delta( p^-_\gamma  - p_X^- )
\end{equation}
which is valid in the high-energy limit where
\begin{equation}
    \frac{p'^- - p^-}{p^-_\gamma} =\xi \frac{  \abs{t}+4M^2}{4 P^+ p_\gamma^-}  \ll 1.
\end{equation}

On the other hand, we can relate the scattering matrix to the scattering amplitude by
\begin{equation}
\begin{split}
    &\mel{X p'}{\hat S -1}{\gamma^* p}
    \\
    =& (2\pi)^4 \delta^{(4)}( p_\gamma + p - p_X - p' ) i \mathcal{M}^{\gamma^* + p \to X + p '}
\end{split}
\end{equation}
and, using Eq.~\eqref{eq:shockwave_scattering}, we can read
\begin{equation}
\label{eq:CGC_scattering}
\begin{split}
   &i \mathcal{M}^{\gamma^* + p \to X + p '}
  =  4  p^-_\gamma \sqrt{p^+ p'^+}
        \sum_{n} 
        \int \dd{[\widetilde \PS]_n}
\\
&
        \times
        2 p_\gamma^- (2\pi) \delta(p_\gamma^- - p_n^-)
\widetilde \Psi_\text{in}^{\gamma^* \to n}
\qty[S^{(n)} - 1]
\qty(\widetilde \Psi_\text{out}^{X \to n})^*.
\end{split}
\end{equation}
Note that often in CGC calculations, the factor $ 4  p^-_\gamma \sqrt{p^+ p'^+}$ (or $ 2  p^-_\gamma \sqrt{p^+ p'^+}$) is absorbed into the definition of the scattering amplitude by defining\footnote{With this rescaling, Eq.~\eqref{eq:CGC_scattering} agrees with Ref.~\cite{Beuf:2024msh} after correcting for an overall minus sign.}:
\begin{equation} 
    \widetilde{\mathcal{M}} = \frac{1}{4 p_\gamma^- \sqrt{ p^+ p^{\prime +}} }\mathcal{M}.
\end{equation}
With this rescaling in mind, we can read the relevant expressions for the CGC scattering amplitudes used in this work from those in the literature.
Note that for the DVCS and DVMP processes considered in this work, where we can approximate $Q^2 \gg p_X^2, \abs{t}$, the rescaling factor can be written as
\begin{equation}
    4 p_\gamma^- \sqrt{ p^+ p^{\prime +}}
    = Q^2 \frac{\sqrt{1-\xi^2}}{\xi},
\end{equation}
where $Q^2 = -p_\gamma^2$ is the photon virtuality.
For inclusive DIS, calculated from the forward scattering amplitude with $X = \gamma^*$ using the optical theorem (see Sec.~\ref{sec:DIS}),
we get
\begin{equation}
 4 p_\gamma^- \sqrt{ p^+ p^{\prime +}}
    = \frac{2 Q^2}{\xbj},
\end{equation}
where $\xbj$ is the Bjorken variable.

The information about the interaction with the target is given by $S^{(n)}$, which can be written in terms of Wilson lines corresponding to the particles in the Fock state $n$.
In the simplest case, corresponding to the quark--antiquark dipole $n= q \bar q$, this scattering is given in terms of the dipole amplitude
\begin{equation}
\begin{split}
    N_{\lambda \lambda'} (\xt,\yt)
    &=
    \frac{1}{\nc} \Tr[1 - S^{(n)}]
    \\
    &=
    1- \frac{1}{\nc} \Tr \expval{V(\xt )V^\dag(\yt)}_{\lambda \lambda'}.
\end{split}
\end{equation}
Here $\xt$ and $\yt$ are the transverse coordinates of the quark and antiquark, $V$ is a Wilson line in the fundamental representation,  and we have written explicitly the helicities $\lambda$ and $\lambda'$ of the incoming and outgoing proton that affect the CGC average $\expval{\ldots}$.
In the CGC framework, the dipole amplitude and other Wilson-line correlators play a similar role to the parton distributions in the collinear factorization, describing the internal structure of the target nucleon.

\section{Parton distributions in the shockwave approximation}
\label{sec:PDF}

Parton distributions can be directly calculated in the shockwave limit by starting from their operator definition, as demonstrated in Ref.~\cite{Bhattacharya:2025fnz}.
We will briefly list the collinear parton distributions that will be used in this work.

The gluon GPD is given by:
\begin{equation}
\label{eq:gluon_GPD}
    \begin{split}
    F^g(\Deltat, x,\xi)
    =&
    \frac{8 \nc  }{\as} 
\sqrt{1 - \xi^2 }
\\
& \times
\int \dd[2]{\kt}
    \widetilde N_{\lambda \lambda'}\qty(
\kt + \frac{1}{2}\Deltat
,-\kt + \frac{1}{2}\Deltat
    ) ,
    \end{split}
\end{equation}
which is written in terms of the Fourier transform of the dipole amplitude:
\begin{equation}
\label{eq:dipole_amplitude_mom}
    \widetilde N_{\lambda \lambda'}(\pt, \qt)
    \equiv 
    \int  \frac{\dd[2]{\xt} \dd[2]{\yt}}{(2\pi)^{4}}
    \frac{e^{i \pt \vdot \xt + i \qt \vdot \yt}}{\abs{\xt -\yt}^2}
     N_{\lambda \lambda'}(\xt, \yt).
\end{equation}
Similarly, the gluon PDF is given by:
    \begin{equation}
    \label{eq:gluon_PDF}
 x f^g(x) =
    \frac{8 \nc }{\as}
\int \dd[2]{\kt}
    \frac{1}{2}
    \sum_\lambda
    \widetilde N_{\lambda \lambda}\qty(
\kt 
,-\kt 
    ) .
    \end{equation}

The parton distributions for quarks are more complicated and require renormalization even at LO.
This follows from the fact that in the CGC, gluon distributions are enhanced by a factor of $1/\as$ compared to the quark distributions, making the gluon contribution to the quark distribution a leading-order effect.
After renormalization, the quark GPD can be written as
\begin{equation}
\label{eq:quark_GPD_final}
\begin{split}
 &  F^q(\Deltat , x ,\xi,\mu_R) 
= \frac{  \sqrt{1-\xi^2}}{\xi}
 \frac{N_c}{2 \pi} 
 \int \dd[2]{\kt} 
 \\
 &\times
 \qty[\widetilde N_{\lambda \lambda'}(\Deltat + \kt,-\kt)
 \hat F^q(\hat \xi_-) 
 +
 \widetilde N_{\lambda \lambda'}( \kt,\Deltat-\kt)
 \hat F^q(\hat \xi_+)],
\end{split}
\end{equation}
where $\hat \xi_\pm = \xi/(x \pm i\varepsilon) $,
\begin{equation}
\begin{split}
    \hat F^q(\hat \xi)
=&
\frac{1}{\hat \xi^2}
\Bigg\{
     \qty[
\log(
\frac{ \mu_R^2}{\kt^2}
)
-\frac{1}{3}
 ]
  \qty[
     2 \hat \xi
     -(1-\hat \xi^2) \log(
     \frac{1+\hat \xi}{1-\hat\xi}
     )
     ]
\\
&+
h(\hat \xi)
\Bigg\}, 
\end{split}
   \end{equation}
and
\begin{equation}
\begin{split}
 h(\hat \xi)
  =&
    \frac{2}{3} \hat \xi
     - (1 -\hat \xi^2 )
     \Biggl[
     2 \qty(
     \li_2(\hat \xi) - \li_2(-\hat \xi)
     )
     \\
     &
        -\frac{5}{3} \log(\frac{1+\hat \xi}{1-\hat \xi})
     +
     \li_2\qty(\frac{1-\hat \xi}{2})
     -\li_2\qty(\frac{1+\hat \xi}{2})
     \\
     &
     +\frac{1}{2}\log^2 \qty(\frac{1-\hat \xi}{2})
     -\frac{1}{2}\log^2 \qty(\frac{1+\hat \xi}{2})
     \Biggr].
\end{split}
\end{equation}
Note that the renormalization scale $\mu_R$ appears here after renormalization in the $\msbar$ scheme.
Similarly, the quark and antiquark PDFs are given by:
\begin{equation}
\label{eq:quark_PDF_renormalized}
    \begin{split}
  & x f^q(x, \mu_R) 
        =
        x f^{\bar q}(x, \mu_R) 
      \\
        =
        & 
 \frac{4 N_c}{3 \pi} 
 \times
 \int \dd[2]{\kt} 
    \frac{1}{2}\sum_\lambda
 \widetilde N_{\lambda \lambda}( - \kt,\kt) 
 \qty[
 \log(
 \frac{\mu_R^2}{\kt^2}) - \frac{1}{3}
 ].
\end{split}
\end{equation}

In these expressions, we can see explicitly the $1/\as$-enhancement of gluon distributions compared to the quark distributions.
This has practical implications for the power counting of the perturbative expansion in the collinear factorization, as this additional factor of $1/\as$ has to be taken into account for a consistent $\as$-expansion of the scattering amplitudes.
This means that quark contributions that would be considered LO in pure collinear factorization need to be treated as of the same order as next-to-leading order (NLO) gluon contributions.

In the pure shockwave limit, one also loses the information about the kinematic constraint $\abs{x} \leq 1$~\cite{Bhattacharya:2025fnz}.
While this might seem problematic at first, in practice
the relevant values in $x$-integrals are dominated by $x \sim \xi$, with $\xi \ll 1$ for the shockwave limit to be valid.
Sensitivity to the region $\abs{x} \gtrsim 1$ would then be an indication of probing physics beyond the validity of the shockwave approximation.

\section{Deeply virtual meson production}
\label{sec:DVMP}

\subsection{Collinear factorization}

Let us first consider DVMP, which turns out to be the simplest of the processes considered in this work.
The reason for this is that the leading-order contribution in collinear factorization comes from both quark and gluon channels, but in the shockwave approximation the quark contribution is suppressed compared to gluons. 
Hence, we only need the gluonic contribution, in which case the leading-twist scattering amplitude for exclusive vector meson production can be written as~\cite{Radyushkin:1996ru,Radyushkin:1997ki,Mankiewicz:1997uy,Diehl:2003ny}:\begin{equation}
\label{eq:DVMP_collinear}
\begin{split}
    &-i \mathcal{M}^{\gamma^*_L + p \to V + p'}
    = -i e \frac{\pi \as}{\nc} \frac{1}{Q}
    \int_0^1 \dd{z}
    f_V \frac{\phi(z)}{z \bar z}
 \\
 &\times
  \int_{-\infty}^\infty \dd{x} \frac{F^g(\Deltat,x,\xi)}{x}
    \qty[
    \frac{1}{\xi-x-i\varepsilon}
    - \frac{1}{\xi+x-i \varepsilon}
    ],
\end{split}
\end{equation}
where $\bar z= 1-z$, and the upper and lower limits of the $x$-integral are taken to infinity due to the lack of the kinematic constraint $\abs{x} \leq 1$ in the shockwave limit.
Here, the distribution amplitude $\phi(z)$ is normalized as
\begin{equation}
    \int_0^1 \dd{z} \phi(z) = 1,
\end{equation}
The decay constant $f_V$ is related to the leptonic width by
\begin{equation}
    \Gamma(V \to l^+ l^-) = \frac{4\pi \aem^2 f_V^2}{3 M_V},
\end{equation}
where $M_V$ is the meson mass.

Substituting our gluon GPD Eq.~\eqref{eq:gluon_GPD} in the shockwave limit to Eq.~\eqref{eq:DVMP_collinear}, we get:
\begin{equation}
\label{eq:DVMP_collinear2}
\begin{split}
    -i \mathcal{M}
    =&
     \frac{\sqrt{1-\xi^2}}{\xi}
   e f_V 
 (2\pi)^2  \frac{4 }{Q }
    \int_0^1 \dd{z}
   \frac{\phi(z)}{z \bar z}
   \\
   &\times
\int \dd[2]{\kt}
    \widetilde N_{\lambda \lambda'}\qty(
\kt + \frac{1}{2}\Deltat
,-\kt + \frac{1}{2}\Deltat
    ) .
\end{split}
\end{equation}
We now wish to compare this expression to the large-$Q^2$ limit of the CGC calculation.

\subsection{Color-glass condensate}

In the high-energy limit,
the amplitude for DVMP can be written as~\cite{Besse:2012ia,Besse:2013muy,Mantysaari:2022bsp}
\begin{equation}
\label{eq:DVMP1}
\begin{split}
 & -i \mathcal{M}^{\gamma^*_L + p \to V + p'}
 =
 Q^2 \frac{\sqrt{1-\xi^2}}{\xi}
 \times
 2e Q f_V 
    \int \dd[2]{\xt} \dd[2]{\yt}
    \\
    &
    \times 
       \int_0^1 \frac{\dd{z} }{4\pi z \bar z}
    e^{i \Deltat \vdot \bt}
    N_{\lambda \lambda'}(\xt,\yt)
    \qty[z \bar z]^{2}  K_0\qty(\sqrt{z \bar z} Q \abs{\rt}  )\phi(z),
\end{split}
\end{equation}
where $\rt = \xt -\yt$ is the dipole size and $\bt = z \xt + \bar z \yt$ is the impact parameter.
To extract the large-$Q^2$ limit from this expression,
it is useful to work in the momentum space.
By Fourier transforming the dipole amplitude with Eq.~\eqref{eq:dipole_amplitude_mom},
we get the expression
\begin{equation}
\label{eq:DVMP2}
\begin{split}
 & -i \mathcal{M}^{\gamma^*_L + p \to V + p'}
 =
 Q^3 \frac{\sqrt{1-\xi^2}}{\xi}
 4e f_V (2\pi)^2
    \int_0^1 \dd{z}  z \bar z  \phi(z)
    \\
    &\times
     \int \dd[2]{\kt} 
    \widetilde N_{\lambda \lambda'}\qty(\kt + z \Deltat , -\kt + \bar z \Deltat)
    \times 
    \frac{z \bar  z Q^2-\kt^2}{(\kt^2 + z \bar z Q^2)^3},
\end{split}
\end{equation}
where we have also used the identity
\begin{equation}
    \label{eq:Kr2_FT}
    \int \dd[2]{\rt} e^{i \rt \vdot \kt} \rt^2 K_0(M\abs{\rt})
    =2\pi \times \frac{4(M^2-k^2)}{(k^2+M^2)^3}.
\end{equation}

We now note that the relevant momentum scale $\kt^2$ in Eq.~\eqref{eq:DVMP2} is determined by the dipole amplitude $\widetilde N$, and hence we expect the transverse momentum to scale as $\kt^2 \sim Q_s^2$, where $Q_s$ is the saturation scale describing gluon saturation effects in the dipole amplitude. 
Expanding in $\kt^2/Q^2$ then corresponds to expanding in twist, such that the leading-twist contribution is given by
\begin{equation}
\label{eq:DVMP_twist_expansion}
      \frac{z \bar  z Q^2-\kt^2}{(\kt^2 + z \bar z Q^2)^3}
      = \frac{1}{\qty[z\bar z Q^2]^2} + \order{\frac{\kt^2}{Q^2}}.
\end{equation}
Using this expansion in Eq.~\eqref{eq:DVMP2} leads to
\begin{equation}
\label{eq:DVMP3}
\begin{split}
 & -i \mathcal{M}^{\gamma^*_L + p \to V + p'}
 =
  \frac{\sqrt{1-\xi^2}}{\xi}
 e f_V (2\pi)^2 \frac{4}{Q}
    \int_0^1 \dd{z} \frac{\phi(z)}{z \bar z} 
    \\
    &\times
     \int \dd[2]{\kt} 
    \widetilde N_{\lambda \lambda'}\qty(\kt + z \Deltat , -\kt + \bar z \Deltat)
    \times \qty[1+ \order{\frac{\kt^2}{Q^2}}].
\end{split}
\end{equation}
Finally, we note that
\begin{equation}
\label{eq:DVMP_dipole_identity}
\begin{split}
      &\int \dd[2]{\kt} 
    \widetilde N_{\lambda \lambda'}\qty(\kt + z \Deltat , -\kt + \bar z \Deltat)
    \\
    =&
      \int \dd[2]{\kt} 
    \widetilde N_{\lambda \lambda'}\qty(\kt + \frac{1}{2} \Deltat , -\kt + \frac{1}{2} \Deltat),
\end{split}
\end{equation}
and thus the leading-twist contribution in the CGC result~\eqref{eq:DVMP3} agrees with Eq.~\eqref{eq:DVMP_collinear2} obtained directly using collinear factorization.

\section{Inclusive deep inelastic scattering}
\label{sec:DIS}

\subsection{Collinear factorization}

Inclusive DIS is generally written in terms of the structure functions $F_2$ and $F_L$, which can be related to the total $\gamma^*+p$ cross section as:
\begin{align}
    F_\lambda(\xbj,Q^2) &\equiv \frac{Q^2}{4\pi^2 \aem} \sigma(\gamma^*_{\lambda_\gamma} + p)
    \\
    F_2(\xbj,Q^2) & \equiv F_L(\xbj,Q^2) + F_T(\xbj,Q^2),
\end{align}
where $\lambda_\gamma = L,T$ is the photon polarization.
By the optical theorem, the cross section can be related to the forward scattering amplitude with:
\begin{equation}
    \sigma(\gamma^*_{\lambda_\gamma} + p) = \frac{\xbj}{Q^2} \Im \mathcal{M}(\gamma^*_{\lambda_\gamma} + p \to \gamma^*_{\lambda_\gamma} + p).
\end{equation}

For $F_2$, only quarks contribute at LO and the gluon contribution only appears at NLO.
However, in the shockwave limit the NLO gluon contribution has the same power counting as the LO quark contribution, and for this reason both of them have to be considered.
The relevant expression in collinear factorization is then given by~\cite{Furmanski:1981cw,Bardeen:1978yd,Ellis:1996mzs}:
\begin{equation}
\begin{split}
    F_2 
    =&
    \sum_f e_f^2 \xbj \qty[ f_q( \xbj, \mu_F ) + f_{\bar q} (\xbj, \mu_F)]
   \\
   &+ \xbj \frac{\as}{2\pi}
    \sum_f e_f^2
    \int_{\xbj}^\infty \frac{\dd{y}}{y} 
2 f_g(y,\mu_F)
\\
&\times
\qty[
P_{qg} \qty(\frac{\xbj}{y})\log \frac{Q^2}{\mu_F^2}
+
C_g\qty(\frac{\xbj}{y}) ] ,
\end{split}  
\end{equation}
where the sum $\sum_f$ goes over (light) quark flavors,
\begin{equation}
    C_g(z)
    = \frac{1}{2}
    \qty{
    \qty[z^2 + (1-z)^2] \log \frac{1-z}{z}
    -1 + 8 z(1-z)
    },
\end{equation}
and 
\begin{equation}
    P_{qg}(z) = \frac{1}{2} \qty[z^2+(1-z)^2].
\end{equation}
Note again that the upper limit of the $y$-integral is taken to infinity due to the lack of the kinematic constraint.
We have also explicitly included the dependence on the factorization scale $\mu_F$, which cancels at the level of the structure function due to the evolution of the parton distributions.

Substituting here our PDFs for the quark~\eqref{eq:quark_PDF_renormalized} and gluon~\eqref{eq:gluon_PDF}, we get:
\begin{equation}
\label{eq:F2_collinear}
  \begin{split}    
    F_2 
    =&
  \frac{8 \nc}{3\pi}   \sum_f e_f^2
    \int \dd[2]{\kt}
 \sum_\lambda  \frac{1}{2} \widetilde N_{\lambda \lambda}\qty(
\kt 
,-\kt 
    ) 
 \qty[
 \log(
 \frac{Q^2}{\kt^2}) + \frac{1}{6}
 ].
\end{split}
  \end{equation}  
Note that the dependence on the factorization scale $\mu_F$ has vanished as expected.
This follows directly from the fact that the parton distributions in Sec.~\ref{sec:PDF} satisfy the collinear evolution equations~\cite{Bhattacharya:2025fnz}.

Similarly, the relevant expression for the $F_L$ structure function is given by~\cite{Cooper-Sarkar:1987cnv,Ellis:1996mzs}:
\begin{equation}
\begin{split}
    F_L
    = 
     \frac{2\as}{\pi}
     \sum_f e_f^2 
    \int_{\xbj}^\infty \frac{\dd{y}}{y}
     \qty(\frac{\xbj}{y})^2
    \qty(1-\frac{\xbj}{y}) y f_g(y,\mu_F),
\end{split}
\end{equation}
where the quark contribution has been neglected due to it being higher order in the shockwave limit.
Substituting our gluon PDF~\eqref{eq:gluon_PDF} here, we get:
\begin{equation}
\label{eq:FL_collinear}
\begin{split}
    F_L 
    =&
     \frac{8\nc}{3\pi}
     \sum_f e_f^2 
\int \dd[2]{\kt}
   \sum_\lambda  \frac{1}{2} \widetilde N_{\lambda \lambda}\qty(
\kt 
,-\kt 
    ) .
\end{split}
\end{equation}

\subsection{Color-glass condensate}

In the CGC approach, it is more convenient to first consider different photon polarizations separately and then calculate the cross section using the optical theorem.
Let us start with the transverse polarization, in which case the scattering amplitude is given by~\cite{Golec-Biernat:1998zce}\footnote{Note that usually the helicity dependence of the dipole amplitude is ignored, and the amplitude is written in terms of $N = \sum_\lambda \frac{1}{2} N_{\lambda \lambda}$ instead.}:
\begin{equation}
\label{eq:DIS1}
\begin{split}
   &- i\mathcal{M}^{\gamma_T^* + p} =\frac{2Q^2}{\xbj} \times Q^2 \frac{\nc}{2} \sum_f \qty(\frac{e e_f}{\pi})^2 \int \dd[2]{\xt} \dd[2]{\yt}
    \\
    &
    \times
       \int_0^1 \frac{\dd{z} }{4\pi z \bar z}
    \sum_\lambda
    \frac{1}{2}
    N_{\lambda \lambda}(\xt,\yt)
    \qty[z \bar z]^2 \qty[z^2 + \bar z^2] K_1\qty(\sqrt{z \bar z} Q \abs{\rt}  )^2. 
\end{split}
   \end{equation}
   To expand in twist,
   we again rewrite this expression in the momentum space.
   By performing the required Fourier transforms, we find:
   \begin{equation}
\label{eq:DIS2}
\begin{split}
 &-   i\mathcal{M}^{\gamma_T^* + p}
    =
    \frac{2Q^2}{\xbj} \times 
    2\pi \nc
  \sum_f  \qty(\frac{e e_f}{\pi})^2 
    \int \dd[2]{\kt} \dd[2]{\kt'}
    \int_0^1\dd{z}
    \\
    &\times 
    \sum_\lambda
    \frac{1}{2}
    \widetilde N_{\lambda \lambda}(\kt , - \kt )
    \frac{ \qty[z \bar z]^2  \qty[z^2 + \bar z^2]  Q^4}{
    \qty[\kt'^2 + z \bar z Q^2]^2
    \qty[(\kt-\kt')^2 + z \bar z Q^2]^2
    }.
\end{split}
\end{equation}
The integral over $\kt'$ can be done using the standard Feynman parametrization, giving us:
\begin{equation}
     \begin{split}
      &
    \int \dd[2]{\kt'}
    \frac{1}{
    \qty[\kt'^2 + z \bar z Q^2]^2
    \qty[(\kt-\kt')^2 + z \bar z Q^2]^2
    }  
    \\
    =&
2\pi
        \int_0^1 \dd{x}
 \frac{1}{
    \qty[x \bar x \kt^2  + z \bar z Q^2]^3
    },   
     \end{split} 
\end{equation}
where we have introduced the Feynman parameter $x$ and denoted $\bar x = 1-x$.
We can now integrate over $z$ and $x$, and find the following expansion:
\begin{equation}
\begin{split}
&
        \int_0^1\dd{z} 
        \int_0^1 \dd{x}
 \frac{ \qty[z \bar z]^2 \qty[z^2 + \bar z^2] x \bar x  Q^6}{
    \qty[x \bar x \kt^2  + z \bar z Q^2]^3
    }   
    \\
    =& 
    \frac{1}{3}
        \qty[
         \log( \frac{Q^2}{\kt^2} )
        - \frac{5}{6}
        +\order{\frac{\kt^2}{Q^2}}
        ].
\end{split}
\end{equation}
Substituting this to Eq.~\eqref{eq:DIS2} gives us the large-$Q^2$ limit:
\begin{equation}
\label{eq:DIS3}
\begin{split}
 -   i\mathcal{M}^{\gamma_T^* + p}
    =&
    \frac{8 \nc }{3 \xbj}
  \sum_f  \qty(e e_f)^2 
    \int \dd[2]{\kt} 
    \sum_\lambda \frac{1}{2}
    \widetilde N_{\lambda \lambda}(\kt, - \kt )
    \\
    &
    \times
        \qty[
         \log( \frac{Q^2}{\kt^2} )
        - \frac{5}{6}
        +\order{\frac{\kt^2}{Q^2}}
        ]
    .
\end{split}
\end{equation}

For longitudinally polarized photons, 
the scattering amplitude is given by~\cite{Golec-Biernat:1998zce}:
\begin{equation}
\label{eq:DIS_L1}
\begin{split}
 &  - i\mathcal{M}^{\gamma_L^* +p}
   =  \frac{2Q^2}{\xbj} \times  Q^2 \frac{\nc}{2}  \sum_f \qty(\frac{e e_f}{\pi})^2 \int \dd[2]{\xt} \dd[2]{\yt}
    \\
    &\times
    \int_0^1 \frac{\dd{z} }{4\pi z \bar z}
    \sum_\lambda \frac{1}{2} N_{\lambda \lambda}(\xt,\yt)
    \times  4 z^3 \bar z^3 K_0 \qty( \sqrt{z \bar z} Q \abs{\rt} )^2,
\end{split}
\end{equation}
and
Fourier transforming Eq.~\eqref{eq:DIS_L1} to the momentum space gives us:
\begin{equation}
\label{eq:DIS_L2}
\begin{split}
 -   i\mathcal{M}^{\gamma_L^*+ p} =&
 \frac{2Q^2}{\xbj} \times      Q^2 \times 8\pi \nc  \sum_f \qty(\frac{e e_f}{\pi})^2 
    \int_0^1 \dd{z} z^2 \bar z^2 
    \\
    &
    \times
    \int \dd[2]{\kt} \dd[2]{\kt'}
     \sum_\lambda \frac{1}{2}\widetilde N_{\lambda \lambda}(\kt,-\kt)
     \\
     &\times
     \frac{z\bar z Q^2 - \kt'^2}{ \qty[\kt'^2 + z \bar z Q^2]^3  }
     \frac{1}{ (\kt'-\kt)^2 + z \bar z Q^2  }.
\end{split}
\end{equation}
The momentum scale $\kt$ is dominated by the saturation scale in the dipole amplitude,
and therefore we can neglect it in the final term:
\begin{equation}
    \frac{1}{(\kt' - \kt)^2 + z \bar z Q^2}
    \approx 
    \frac{1}{\kt'^2 + z \bar z Q^2} + \order{\frac{\kt^2}{Q^2}}.
\end{equation}
Using this approximation, the remaining integrals over $\kt'$ and $z$ can be evaluated analytically, giving us:
\begin{equation}
\label{eq:DIS_L3}
\begin{split}
 -   i\mathcal{M}^{\gamma_L^*+ p}
       =&  \frac{8 \nc}{3 \xbj}    \sum_f  \qty(e e_f)^2 
    \int \dd[2]{\kt}
      \sum_\lambda \frac{1}{2}\widetilde N_{\lambda \lambda}(\kt,-\kt)
    \\
    &
    + \order{\frac{\kt^2}{Q^2}}.
\end{split}
\end{equation}

Substituting now the scattering amplitudes for the transversely~\eqref{eq:DIS3} and longitudinally~\eqref{eq:DIS_L3} polarized photons, we get exact agreement with the structure functions $F_2$~\eqref{eq:F2_collinear} and $F_L$~\eqref{eq:FL_collinear} calculated using collinear factorization.
This includes not only the logarithmically enhanced terms but also the constant terms, verifying the consistency between the two approaches.

\section{Deeply virtual Compton scattering}
\label{sec:DVCS}

\subsection{Collinear factorization}

The dominant contribution to DVCS comes from the case where both the initial and final photon have the same transverse polarization.
In this case, the scattering amplitude in collinear factorization is given by~\cite{Ji:1997nk,Ji:1998xh,Mankiewicz:1997bk,Schonleber:2023ouo}:
\begin{equation}
\begin{split}
 & -i \mathcal{M}^{\gamma^*_{\pm 1} + p \to \gamma_{\pm 1} + p'}  
 \\
     =&-i e^2
     \int_{-\infty}^\infty \frac{\dd{x}}{\xi}
     \Biggl[  \sum_q F^q(\Deltat,x,\xi; \mu_F) C_0^{S,q}\qty(z,\mu_F)
     \\
     &+
     \frac{1}{\xi}
     F^g(\Deltat,x,\xi;\mu_F) C_1^{S,g}\qty(z,\mu_F)
     \Biggr]
\end{split}
\end{equation}
where again we include the LO quark and NLO gluon contributions that are of the same order in $\as$ in the shockwave limit.
The coefficient functions are given by
\begin{equation}
    C_0^{S,q}(z,\mu_F)=\frac{e_q^2}{2} \qty[\frac{1}{z}- \frac{1}{\bar z}]
\end{equation}
and
\begin{equation}
\begin{split}
    C_1^{S,g}(z,\mu_F)=&
    \sum_f e_f^2 \frac{\as}{4\pi} \frac{1}{4 z \bar z} 
    \Biggl[
    \log(\frac{\mu_F^2}{Q^2}) \frac{z}{\bar z} \log z
    \\
    &
    - \frac{z}{2\bar z} \log^2 z
    + \frac{1+z}{\bar z} \log z
    + (z \leftrightarrow \bar z)
    \Biggr],
\end{split}
\end{equation}
where $z = \frac{1}{2}\qty(1- \frac{x}{\xi-i\varepsilon})$ and $\bar z = 1-z$.
Here, the skewness variable is shifted to $\xi -i \varepsilon$ to keep track of the poles in the imaginary plane when integrating over $x$, and in the end we can take $\varepsilon \to 0$.

Substituting our quark GPD~\eqref{eq:quark_GPD_final}, we see that we need to evaluate the following integrals:
\begin{align}
\begin{split}
     \int_{-\infty}^\infty \frac{\dd{x}}{\xi}\hat F^q(\hat \xi_-)  C_0^{S,q}(z, Q)
    =& 2 \pi i e_q^2 \hat F^q(1+i\varepsilon) 
    \\
    =&4\pi i e_q^2
\log(
\frac{\mu_F^2}{\kt^2}
)
\end{split}
    \\
    \begin{split}
        \int_{-\infty}^\infty \frac{\dd{x}}{\xi}\hat F^q(\hat \xi_+)  C_0^{S,q}(z, Q)
    =& -  2 \pi i  e_q^2 \hat F^q(-1-i\varepsilon) 
 \\
 =&4\pi i  e_q^2
\log(
\frac{\mu_F^2}{\kt^2}
)
    \end{split}
\end{align}
which can be done using the residue theorem.
The quark contribution is then given by:
\begin{equation}
\begin{split}
     &   \int_{-\infty}^\infty \frac{\dd{x}}{\xi}
     F^q(\Deltat,x,\xi; Q^2) C_0^{S,q}(z,Q)
     \\
     =&
      \frac{  \sqrt{1-\xi^2}}{\xi}
2i \nc  e_q^2
 \int \dd[2]{\kt} 
 \log(
\frac{\mu_F^2}{\kt^2})
 \\
 &\times 
 \qty[\widetilde N_{\lambda \lambda'}(\Deltat + \kt,-\kt)
 +
 \widetilde N_{\lambda \lambda'}( \kt,\Deltat-\kt)]
  .
\end{split}
\end{equation}

For the gluon contribution, we have the following integral:
\begin{equation}
\begin{split}    
       &  \int_{-\infty}^\infty \frac{\dd{x}}{\xi^2}
 C_1^{S,g}\qty(z,Q)
 = -\frac{i \as}{\xi}  \sum_f e_f^2 
 \qty[ 1 + \frac{1}{2} \log( \frac{\mu_F^2}{Q^2}) ],
\end{split}
\end{equation}
which can again be evaluated using the residue theorem.
Finally, we can write the DVCS amplitude as:
\begin{equation}
\label{eq:DVCS_collinear}
\begin{split}
 & -i \mathcal{M}^{\gamma^*_{\pm 1} + p \to \gamma_{\pm 1} + p'}  
    =   2\nc  \sum_f (ee_f)^2  
      \frac{  \sqrt{1-\xi^2}}{\xi}
      \int \dd[2]{\kt} 
      \\
      &
      \times
       \qty[ \log(
\frac{Q^2}{\kt^2})
 - 2
]
 \qty[\widetilde N_{\lambda \lambda'}(\Deltat + \kt,-\kt)
 +
 \widetilde N_{\lambda \lambda'}( \kt,\Deltat-\kt)].
\end{split}
\end{equation}
We see that the dependence on the factorization scale $\mu_F$ cancels identically, as in the case of inclusive DIS in Sec.~\ref{sec:DIS}.

\subsection{Color-glass condensate}

In the CGC framework, the amplitude is given by~\cite{Bartels:2003yj,Kowalski:2006hc,Hatta:2017cte}:
\begin{equation}
\label{eq:DVCS1}
\begin{split}
  &-i \mathcal{M}^{\gamma^*_{\pm 1} + p \to \gamma_{\pm 1} + p'}  
   =
   Q^2 \frac{\sqrt{1-\xi^2}}{\xi}
   \times 
   \frac{\nc}{2}  \sum_f \qty(\frac{e e_f}{\pi})^2
    \\
    &
    \times
     \int \dd[2]{\xt} \dd[2]{\yt}
    \int_0^1 \frac{\dd{z} }{4\pi z \bar z}
    e^{i \Deltat \vdot \bt}
    N_{\lambda \lambda'}(\xt,\yt) \qty[z \bar z]^{3/2}\\
    &\times \qty[z^2 + \bar z^2] \frac{Q}{\abs{\rt}} K_1\qty(\sqrt{z \bar z} Q \abs{\rt}  ),
\end{split}
   \end{equation}
   which in the momentum space can be written as
\begin{equation}
\label{eq:DVCS2}
\begin{split}
   - i\mathcal{M}
    =&
   Q^2 \frac{\sqrt{1-\xi^2}}{\xi}
   \times 
   2 \nc   \sum_f  \qty(e e_f)^2 \int \dd[2]{\kt} 
    \int_0^1 \dd{z} 
    \\
    &
    \times
    \widetilde N_{\lambda \lambda'}\qty(\kt + z \Deltat , -\kt + \bar z \Deltat)
    \times 
    \frac{  \qty[z^2 + \bar z^2]   z \bar z Q^2}{(\kt^2 + z \bar z Q^2)^2}.
\end{split}
\end{equation}
To leading order in the large-$Q^2$ expansion, we note that we are allowed to substitute
\begin{equation}
\label{eq:N_substitution}
\begin{split}
      &\widetilde N_{\lambda \lambda'}\qty(\kt + z \Deltat , -\kt + \bar z \Deltat)
    \times \qty[z^2 + \bar z^2] 
    \\
    \Rightarrow &
 \widetilde N_{\lambda \lambda'}\qty(\kt + \Deltat , -\kt ) \times z^2
 + \widetilde N_{\lambda \lambda'}\qty(\kt  , -\kt +  \Deltat) \times \bar z^2
\end{split}
\end{equation}
in the $z$-integral.
This is because in the large-$Q^2$ limit, the main contribution comes from the endpoints $z\to 0$ and $z \to 1$ of the $z$-integral.
With this substitution, we can calculate the $z$-integral and get the following expansion:
\begin{equation}
\label{eq:DVCS_z-integral}
\begin{split}
    &\int_0^1 \dd{z} \frac{z \bar z \times z^2 Q^4}{(\kt^2 + z \bar z Q^2)^2}
    =
    \int_0^1 \dd{z} \frac{z \bar z \times \bar z^2Q^4}{(\kt^2 + z \bar z Q^2)^2}
    \\
    =&  \log\qty(\frac{Q^2}{\kt^2}) - 2 + \order{\frac{\kt^2}{Q^2}} .
\end{split}
\end{equation}
Using Eqs.~\eqref{eq:N_substitution} and~\eqref{eq:DVCS_z-integral}, the DVCS amplitude can be written as
\begin{equation}
\label{eq:DVCS3}
\begin{split}
   - i\mathcal{M}
    =&
    \frac{\sqrt{1-\xi^2}}{\xi}
   \times 
   2 \nc   \sum_f  \qty(e e_f)^2 \int \dd[2]{\kt} 
    \int_0^1 \dd{z} 
    \\
    &
    \times
    \qty[
     \widetilde N_{\lambda \lambda'}\qty(\kt + \Deltat , -\kt )
     +
      \widetilde N_{\lambda \lambda'}\qty(\kt , -\kt +  \Deltat)
    ]
    \\
    &
    \times 
  \qty[ \log\qty(\frac{Q^2}{\kt^2}) - 2 + \order{\frac{\kt^2}{Q^2}}],
\end{split}
\end{equation}
which is found to be in exact agreement with the collinear limit~\eqref{eq:DVCS_collinear}, including both the logarithmic and the constant terms.

\section{Conclusion}
\label{sec:conclusion}

In this work, we have considered inclusive and exclusive processes in deep inelastic scattering from two different approaches.
In the first approach, we use the expressions for scattering amplitudes from collinear factorization, written in terms of twist-2 parton distributions, and substitute the expressions for the parton distributions obtained with the shockwave approximation.
In the second approach, we start from the high-energy factorization where the scattering amplitudes are written in terms of Wilson-line correlators, and then expand in twist.
By considering specifically DVMP, DVCS, and inclusive DIS, we have shown that both approaches yield the same result in the limit of high energy and large $Q^2$.
This includes both the logarithmically enhanced contributions, related to the evolution of the parton distributions, and the finite terms.
Notably, a crucial part of showing the equivalence between the different approaches is the absence of the kinematic constraint $\abs{x} \leq 1$, in line with the shockwave approximation.
This shows the consistency between collinear and high-energy factorizations at LO, suggesting a possibility for a combined approach in this complementary region of the two factorization approaches.

Our results also suggest a way for including  the resummation of large logarithms of momentum scales within the CGC framework, by isolating the logarithmically enhanced contribution. 
In general, these logarithms could ruin the perturbative expansion for large $Q^2$, as they scale as $\log (Q^2/ \kt^2) \sim \log (Q^2/ Q_s^2)$, with the saturation scale $Q_s$ being quite modest ($\lesssim \SI{1}{GeV}$) at the energies of the current experiments.
As we have demonstrated, these contributions can be understood as coming from parton distributions in the shockwave limit, and as such they can be resummed by the appropriate collinear evolution equations of the parton distributions.
Such resummations are expected to improve the accuracy of phenomenological comparisons to the experimental data for large  photon virtualities $Q^2 \gg Q_s^2$.

The natural next step is to extend this matching to NLO, where it will be possible to explicitly see the interplay between the high-energy evolution (such as the BK and JIMWLK equations) and the collinear evolution (i.e. DGLAP and the GPD evolution equations).
This will require extending the calculation of the parton distributions in the shockwave approximation to NLO, and for this reason it is left for future work.

Our procedure also suggests a clean way to calculate the leading-twist terms in the high-energy limit, by first writing the relevant parton distributions in the shockwave approximation and then using the appropriate expressions from the collinear or TMD factorization.
While extracting the leading-twist contribution from the LO CGC expressions is fairly straightforward, the situation becomes much more complicated at NLO (see e.g. Refs.~\cite{Caucal:2022ulg,Caucal:2023fsf,Caucal:2023nci,Caucal:2024nsb,Hauksson:2024bvv,Kaushik:2025roa}).
Following the method of this work, the most difficult part is calculating the parton distributions at higher orders; however, due to their universality, the same parton distributions can be used to study a wide range of different processes.
We thus expect that calculating directly the parton distributions in terms of the CGC Wilson lines can provide an easier way for calculating higher-order contributions in the kinematic region of high energy and large momentum scales.

\section{Acknowledgments}
Z.K., D.P., and J.P. are supported National Science Foundation under grant No.~PHY-2515057. This work is also supported by the U.S. Department of Energy, Office of Science, Office of Nuclear Physics, within the framework of the Saturated Glue (SURGE) Topical Theory Collaboration.

\bibliographystyle{JHEP-2modlong.bst}
\bibliography{references.bib}

\end{document}